\documentclass[twoside,12pt]{article}
\usepackage{cite}      

\begin{document}

\title{Direct Observation Limits  \\ 
on Antimatter Gravitation \\
}

\author{
Mark Fischler\footnote{Fermi National Accelerator Laboratory}, 
Joe Lykken\footnotemark[1], and 
Tom Roberts\footnote{Muons, Inc. and Fermi National Accelerator Laboratory}
}
\date{May 20, 2008}
\maketitle

\abstract

The proposed Antihydrogen Gravity experiment at Fermilab (P981) will  
directly  measure  the gravitational
attraction $\bar{g}$ between antihydrogen and the Earth,
with an accuracy of 1\% or better.   
The following key question has been
asked by the PAC:

\begin{quote}
Is a possible 1\% difference between $\bar{g}$ and $g$ already ruled out by
other evidence? 
\end{quote}

This memo presents the key points of existing evidence, 
to answer whether such a difference is ruled out (a)
on the basis of direct observational evidence; and/or (b) 
on the basis of indirect evidence, combined with reasoning based on 
strongly held theoretical assumptions.
The bottom line is that there are no direct observations 
or measurements  of
gravitational asymmetry which address the antimatter sector.  
There is evidence which by
indirect  reasoning can be taken to rule out such a difference, but the 
analysis needed to draw that conclusion rests on models and assumptions 
which 
are in question for other reasons and are thus 
worth testing.  
There is no compelling evidence or 
theoretical reason to rule out such a difference at the 1\% level.  

\newpage
\tableofcontents
\newpage

\section{The Nature of Evidence About Gravitational Asymmetry}

P981 is a proposed Fermilab experiment\cite{loi} to measure, 
using atomic beam interferometry,
the gravitational acceleration of antihydrogen.
The experimenters are confident that 1\% accuracy or better 
would be achieved.  
During an initial presentation to the PAC, 
the following question was posed:
\begin{quote}
Is a possible 1\% difference between $\bar{g}$ and $g$ already ruled out by
other evidence? 
\end{quote}

We take this as asking whether a difference of that magnitude
is ruled out either by direct observation,
or by experimental results that lead to incontrovertible (though indirect)
evidence. 
If experimental results exist that cannot be
reconciled with this level of gravitational asymmetry,
assuming any sensible theory,
then one could say the asymmetry is ruled out.
We also consider whether any known theoretical consideration 
(based, for instance, on some gedanken experiment) irrefutably
rules out such an effect.

Of course, one might  
simply   
assert that the only ``sensible'' theory 
at low energies and macroscopic distances is 
standard General Relativity with the Weak Equivalence Principle holding 
exactly,
and further assert that given this framework, such an asymmetry is 
impossible. 
There are three major weaknesses of this ``proof by assertion'':
\begin{itemize}
\item 
There is evidence (from the accelerating expansion of the universe ) that 
calls into question either 
the composition of the universe or the correct theory to describe the
interactions driving that expansion.  It is not out of the question that
a revised theory correctly describing the expansion could include interactions
that distinguish antimatter from matter.
\item 
The simplest theoretical model that predicts a non-null
result for the antihydrogen experiment postulates 
new vector and scalar mediated forces that couple to some combination
of baryon and lepton number (or any other quantity that distinguishes matter
from antimatter).  
If the vector and scalar forces are of equal magnitude, their effects will
to some high accuracy cancel for ordinary matter.
Nieto and Goldman\cite{nieto} (section 8) point out that in 
this context, ``the more precisely anomalous gravitational effects are
ruled out in earth-based matter-matter experiments, the more unrestricted
is the possibility that there can be a significant anomalous gravitational
acceleration of antimatter.''
These ``fifth forces'' do not
invalidate General Relativity any more than do the gauge forces
of the Standard Model.
\item 
One must always be careful about hidden assumptions. 
One obviously relevant 
hidden assumption is CPT symmetry. While this can
be proven rigorously for asymptotically flat-space relativistic 
local field theories, string theory calls into question both locality and
the fundamental existence of an S-matrix.
\end{itemize}

We will discuss in section \ref{obstacles} whether there is general
agreement among the gravitational theory community
that antimatter gravitational asymmetry can or cannot be accommodated by a
consistent theory.  First, however, we will examine consequences of 
existing down-to-earth measurements and observations directly or indirectly
bearing on this question.

\subsection{Meaning of ``Direct Observation''}

What constitutes a ``direct observation'' of the value of some 
physical quantity ($\bar{g}/g$ in this case)?  We want to avoid 
general philosophic discussions, so we will define a ``direct observation''
of a value as evidence from which the value 
can be deduced without depending on any assumptions which a reasonable 
and expert scientist might call into question.
This implies that the analysis can be done in a straightforward and essentially 
model-independent manner.

In the case at hand, we are asked to address whether  
$\bar{g}/g \neq 1$ is ruled out at the 1\% level by existing evidence.
 
\subsection{Categories of (Experimental) 
	Evidence Against Gravitational Asymmetry}

Although there are multiple experiments measuring asymmetry between one sort
of particle and another, in relation to gravity, they all fall into three
general classes.
First, 
there are experiments that drop, throw, or (gravitationally or otherwise)
deflect different types of matter and measure the resulting paths or forces
\cite{rpp}\cite{will}\cite{frank}\cite{adelberger1}\cite{nesv}. 
These can be considered to be ``direct'' measurements, since they address 
the quantity (force or acceleration) in which we are specifically interested,
and depend on no framework (other than that needed to arrive at the force or 
acceleration results themselves) to interpret their data.
We can think of this as ``Newtonian'' evidence, since such results can
bear on the issue without any framework beyond $F = m a$.
Direct observations are discussed in section \ref{direct}.

Second,
there are experiments (or deductions made incidental to other experiments)
for which analysis of certain negative results allow
one to ``rule out'' gravitational asymmetry at some level, by applying 
some properties of gravity
\cite{good}\cite{kenyon}\cite{hughes}\cite{gabrielese}. 
The crucial property in these experiments involves gravity rescaling
a particle's observed time relative to lab time.
These can be considered to be ``indirect'' measurements, 
since one must use at least some generic property of the underlying
gravitational theory to derive the asymmetry limits from their data.
We can think of this as ``Einsteinian'' evidence,
since such results impact the issue of gravitational asymmetry via
concepts that are not present in non-relativistic physics.
These deductions are discussed in section \ref{timescale}.

A third group of arguments is based on astrophysical evidence about 
neutrinos and antineutrinos: 
observation of equal time-of-flight
for neutrinos and antineutrinos originating in  
supernova SN1987A\cite{losecco}, 
and inferences from solar\cite{minakata} or terrestrial\cite{gasperini}
neutrino oscillation observations.
We will see in section \ref{neutrino} that these arguments are uniformly
weaker evidence against the possibility of antimatter gravitational asymmetry
than the ``Einsteinian'' evidence.

\subsection{What Fraction of the Antiproton Mass Is Relevant?} 

An observation has  
been made\cite{kayser}  that even if $\bar{g}/g = -1$,
the antiproton would not ``fall up'' with acceleration $g$.  
This is because a large fraction of the inertial mass of a proton or 
antiproton comes from its binding energy (the gluon field).  
While it is unclear whether to use the low momentum current-quark masses 
or some other
quantity (perhaps a chiral mass) as the component which distinguishes between
matter and antimatter, 
if we say that the current-quark mass is critical, then the mass 
in question is only about 1\% of the overall antiproton mass.
Taking that viewpoint, the 1\% level represents a {\em maximum}
plausible effect.

We cannot refute that viewpoint, but it does depend
on the Standard Model, testing of which has been the mission of Labs
world-wide for several decades.
Validation of predictions of the Standard Model 
in circumstances involving gravity 
is certainly fair game for investigation.

We note that the same 1\% argument (that most of the antiproton mass
consists of ``ordinary'' binding energy) applies to the analysis of all the
experimental evidence we will be looking at below, and weakens every statement
the experiments make about limits on $\bar{g}/g$.  
However, the most precise experimental evidence of each type sets a limit
(if accepted as conclusively bearing on $\bar{g}-g$) which is much smaller than
1\%.

The bottom line about this ``quark fraction'' reasoning is that it is moot to
the issue of whether a 1\% effect can possibly be present in antihydrogen 
measurements:  If either class of experimental results is accepted as placing
irrefutable or direct limits on $\bar{g}$, then the proposed experiment cannot 
detect an effect, even ignoring the ``quark fraction'' reasoning.  And if
none of that evidence presents a conclusive measurement, then there is 
potential for an effect as large or larger than the 1\% level.

\section{Limits Set By Force or Acceleration Observations}
\label{direct}

A $\bar{p}$ trajectory experiment sensitive to gravity would provide 
a direct  measurement of the value of $\bar{g}$.  This sort of experiment 
was proposed at Fermilab\cite{phillips} and suggested as part of an
experiment at LEAR\cite{lear}, but measurement of the trajectory
of (charged) antiprotons
has never been done with sufficient sensitivity to observe gravitational 
acceleration.
There have been neutron 
free-fall experiments\cite{frank}\cite{nesv}, 
which verify that neutrons have the same value of $g$ as 
ordinary neutron-proton matter, to an accuracy of .2\%.  
Again, 
such an experiment with antineutrons would constitute a direct observation 
of $\bar{g}/g$, but that hasn't been done either, and is not likely to be
feasible because antineutrons can't easily be cooled -- any cooling mechanism
involving matter interaction turns into an annihilation mechanism instead.
The most sensitive neutron experiment that has been performed involved 
bouncing neutrons on a matter surface, and thus is inherently impossible 
to replicate with antineutrons.   

E\"{o}tv\"{o}s-style experiments\cite{adelberger1}, 
measuring the gravitational  
attraction between a heavy mass and test masses of
various element composition, 
directly measure the difference in coupling
to elementary particle mass and to binding energy.  For example, 
the binding energy fraction for beryllium is half a percent less than that
for copper, and the experimental limit on 
$|\Delta a/a|$ for beryllium versus copper
is $2.5\cdot 10^{-12}$\cite{will}\cite{rpp}.  
This puts a limit on the difference of gravitational
attraction for ordinary proton-neutron mass to nuclear binding energy, of
about $5\cdot 10^{-10}$, and this has to be considered as direct observation,
based on a force measurement.

At any rate, neither the E\"{o}tv\"{o}s-style 
experiments nor the neutron
free-fall experiments 
make any statement about antimatter,
unless one stretches to say that different elements have different virtual 
antimatter proportions.  
The argument would go as follows:
Suppose some fraction of the mass of the proton is attributable to 
masses of the (virtual)
anti-quarks in the ``sea.''  
Then there will be 
analogous 
antimatter fractions in various nuclei.
These antimatter fractions would be expected to have about the 
same value as that for the proton,
but are unlikely to match precisely.
Therefore, the elements tested will likely have 
antimatter fractions which differ by some (second-order small) amount.
If this difference is as large as one part in a million,
then the E\"{o}tv\"{o}s-style
experiments set a limit on antimatter gravitation asymmetry at about 
$5\cdot 10^{-3}$. 

This argument is greatly weakened by the cosmological constant problem.
If gravity couples to virtual particles and antiparticles inside nucleons,
it presumably also couples to virtual particles and antiparticles that
constitute vacuum energy. This leads to the famous estimate of the cosmological
constant that disagrees with observation by (at least) 120 orders of magnitude.
Obviously there is something fundamentally wrong with our understanding
of how to relate gravity to virtual effects.
 
Furthermore, even if the naive picture of gravity coupling to virtual
antiparticles were correct,
there is no strong justification for saying
that the antimatter content fraction differs between one element and another,
and certainly no estimate has been derived of the size of this putative effect. 

So, although direct acceleration observations do fix gravitational indifference
to the nature of mass (among ordinary matter and binding energy) 
at much better than the 1\% level, they do not 
constitute direct observations of this property applied to antimatter.

\subsection{Combining E\"{o}tv\"{o}s-style measurements with inferences from
other observations}

A stronger line of reasoning combines the exquisite precision of 
E\"{o}tv\"{o}s-style ordinary-matter measurements with results
from other gravitation experiments and assumptions about the forms
of possible equivalence-principle-violating interactions, to infer
limits on the antimatter sector from those results.
Detailed by the E\"{o}t-Wash collaboration in section F of \cite{adelberger1},
and expounded more succintly in \cite{ericson} and \cite{adelberger2}, 
the argument is:
\begin{enumerate}
\item
Assume deviation from the weak equivalence principle takes the form of 
vector and scalar gravity-like forces, where the vector-like force may
affect antimatter differently from matter.
\item
The E\"{o}t-Wash results, by comparing elements with different proton-neutron
fractions, force very precise cancellation between these gravi-scalar and 
gravi-vector forces in the ordinary matter sector.  
\item  
Dimensional considerations restrict the nature of the vector and scalar forces
to forms containing a limited set of parameters.  The cancellation in the
E\"{o}tv\"{o}s-style experiments strongly constrains the relative values of 
some of these parameters.  
\item 
Incorporation of the observational limit on difference between the proton and
anti-proton $|q/m|$, 
and of Laser Geodynamics data that place a limit on deviation
from $1/r^2$ behavior, then places limits in other realms.
These over-constrain the parameters unless the
``cancelling'' scalar and vector forces are very small.
\end{enumerate} 

This reasoning leads to a limit on the value of antihydrogen gravitational
asymmetry, at the $10^{-5}$ level.  

However, Goldman {\it et al.}\cite{goldman}
points out a weakness in the third point of the above argument:
The restrictions on possible forms for the scalar and vector interactions
stem from dimensional constraints which are applicable to renormalizable
theories.  These do not generally apply to gravitation theories.  
Loosening these restrictions leaves open the possibility of near-precise 
cancellation for ordinary matter, enforced by some approximate symmetry.
And to the extent that such cancellations are possible, 
the E\"{o}tv\"{o}s-style experiments do not 
constitute direct observations of gravitational indifference
to the nature of mass where antimatter is concerned.

\section{Limits Based on Time-Scaling Effects}
\label{timescale}

There are two experiment-based arguments that put sensitive limits on 
matter-antimatter gravitational asymmetry:  
the (tiny value of the) $K^0 - \overline{K^0}$ 
oscillation rate \cite{good} \cite{kenyon} \cite{cplear}, 
and the equality of cyclotron frequencies for  
the proton and the antiproton \cite{hughes}\cite{gabrielese}).  
  
We discuss these together because they both hinge on the general-relativistic
effect of gravitational redshift.  By this, we mean the effect that 
to an observer at the top of some gravitational potential, clocks on objects 
at a lower potential appear to tick more slowly.  A consequence
of the Weak Equivalence Principle (objects travel along geodesics), it also
has been verified by direct observation of redshifts (e.g. downward
traveling photons and shifts of solar spectral lines;
see table 2.3 of reference \cite{will})
matching the predicted
value.  

The cyclotron frequency evidence is particularly on-point for antihydrogen.
This is discussed by Hughes and Holzscheiter\cite{hughes}.
The idea is that the cyclotron frequency for a proton depends on the  
proton's electromagnetic properties (mostly charge, but at higher orders
perhaps magnetic moment) and mass.  The  cyclotron frequency is a rest-frame
quantity.  Since the mass and absolute charge of a $\bar{p}$ are identical
to those of a $p$, the inherent cyclotron frequency will be identical.  
However, the quantity that external measurers will observe as the 
cyclotron frequency might 
not be identical if the time unit for a $\bar{p}$ is different from that
for a $p$!  

Now let's place the ${p}$ (or $\bar{p}$) into a gravitational field as follows:
The metric is asymptotically flat, but contains a concentration of mass
centered at some distance $R$ from the proton.  Relative to an observer near
the concentration of mass, the local time of the proton is sped up by a factor 
of $1+ GM/Rc^2$.  That is, there is a time acceleration effect proportional to
the ``absolute gravitational potential'' felt by the proton.

Next, assume the $\bar{p}$ interacts differently with that gravitational 
field.  Its time acceleration factor will instead be $1+ (\bar{g}/g)(GM/Rc^2)$.
Thus its cyclotron frequency, as observed by ``we the accelerator people''
(who being made of ordinary matter will share the acceleration of the proton)
will be different by an amount proportional to $(1-\bar{g}/g)(GM/Rc^2)$.
Such a difference would (in the experiments described by 
Gabrielese\cite{gabrielese}) manifest as a difference in  
charge-to-mass ratio $|q/m|$ for the proton and antiproton.

What should we use for $R$ and $M$ if such an argument is true?  
Well, to a Good approximation (pun intentional\cite{good} ), the local galactic
super cluster sits isolated in an asymptotically flat space, and produces
a fairly large absolute gravitational potential 
($\phi_g = GM/rc^2 \approx 3 \times 10^{-5}$)\cite{kenyon}.
Since we know the proton and antiproton have the same cyclotron
frequency to a part in $10^{10}$\cite{gabrielese}, 
this sets an upper bound on 
$|1-\bar{g}/g|$ of about $5 \times 10^{-4}$.  

Similarly, the observation of near-perfect non-regeneration of $K^0$ 
(and thus decaying $K_S$'s) from the $K_L$ state (the small
regeneration is observation of CP violation) 
would be utterly destroyed by differences in the local time steps between the
$K^0$ and the $\overline{K^0}$.  
A simple exposition of the argument appears in section 7.1 of 
Nieto and Goldman\cite{nieto}.  
If the gravitational potential 
$\phi_g$ is assumed to have absolute meaning, then the energy difference
between $K^0$ and $\overline{K}{}^0$ due to 
antimatter gravitational asymmetry
should be $\frac{g-\bar{g}}{g} M_K \phi_g$,
which for 
hypothetical ``antigravity'' 
($\bar{g} = - g$) is
$2 M_K \phi_g$.  
Since $K_L$ is a superposition of 
$K^0$ and $\overline{K}{}^0$, any $K_L$ in a beam should regenerate some $K_S$ 
because the relative time variation of the 
$K^0$ and $\overline{K}{}^0$ components
is $\exp(2 i M_k \phi_g t / \hbar)$. 
Now regeneration of $K_S$ is observed as CP violation, and we know the
size of that apparent CP violation.   
Tying all this together, we get, for some given $\phi_g$, a limit on 
$\bar{g} - g$.  

Using the same time effect that we used in the cyclotron frequency argument,
of $(1-\bar{g}/g)(GM/Rc^2)$, you can infer that $|1-\bar{g}/g|$ applied to the 
$\overline{K^0}$ versus the $K^0$ is at most two parts in a billion
\cite {cplear}\cite{hughes}.  
However,
unlike the cyclotron frequency observation, this speaks mostly 
to the gravitational attraction on $s$ and $\bar{s}$ quarks; it is a
bit of a leap to say that this proves the issue for matter versus antimatter
in general.

What we have in both cases is evidence which hinges on the cosmological
argument.  That, in turn, hinges on:
\begin{itemize}
\item CPT invariance 
\item the notion of a meaningful absolute gravitational potential
\item the idea that the appropriate
potential to use is that of the local super-cluster
\end{itemize}

While CPT invariance is rock-solid provable assuming only causality, 
local Lorentz invariance, and asymptotically flat space-time, it is 
in fact possible that the doubt about the flatness assumption 
(which may not be true on a cosmological scale) raises doubt about the 
applicability of CPT to our analyses.  Kenyon\cite{kenyon}, for example, 
couches his analysis in terms of a CPT-violating mass difference between the 
$K^0$ and $\overline{K}{}^0$.  

Nieto and Goldman\cite{nieto}
question the use of an absolute gravitational potential,
upon which rest the conclusions from both the cyclotron frequency and the
$K^0 - \overline{K^0}$ observations.  For an infinite-range force, 
what matters are 
potential {\em differences}.  
They present a formalism that recomputes the effects using potential 
differences.  Using this approach, Chardin and Rax\cite{chardin} 
derives that the observed CP violation is just about the right magnitude to be 
alternatively explained by antigravity, i.e., $\bar{g}/g= -1$. 

In the framework of the geometric picture presented by 
General Relativity, the simple ``absolute potential'' calculation 
is justified.  
But if there is antimatter gravitational asymmetry, 
then General Relativity is violated at least in the antimatter sector;
the overall geometric picture and thus the  
absolute potential argument come into question.  
That is, in order to use the absolute potential based arguments to
interpret these results as limiting possible antimatter asymmetry, 
one must logically start with the premise that General Relativity holds
in its particulars, and thus that antimatter asymmetry cannot be present
{\it a priori}.  

Even assuming the validity of the use of an 
absolute gravitational potential, one must decide which system
defines that potential.  
That in turn depends on what one assumes about the nature (length scale)
of any possible matter-antimatter-asymmetric gravity-like force.  
Kenyon selected the definable system giving the largest absolute potential;
this is natural if one wants to avoid any long but finite length scales.
But this is not unavoidable.  Good's original paper\cite{good}, for example, 
chose the gravitational potential due to the Earth; this pushes 
the cyclotron frequency limit on asymmetry above the 1\% level
(though the kaon system limit remains below 1\%).   Nieto and
Goldman
present models involving finite-range vector and scalar gravity fields,
in which apparent violation of $\bar{g}=g$ could easily be seen in the
envisioned experiment.

Ultimately, while 
we consider this class of observations (particularly the cyclotron frequency
equality) to be indirect
evidence of the equality of $g$ and $\bar{g}$ to high accuracy, 
we cannot consider this to be a direct 
observation of their ratio, nor an irrefutable indication that this 
equality must hold, because there are disputable steps in the analysis
going from the observations to those conclusions.
This interpretation is supported by the
statements of Chardin and Rax, Nieto and Goldman, and also in section 5.4 of
Will\cite{will}.

\section{Inferences From Astronomical (Anti)Neutrino Observations}
\label{neutrino}

Several observations from the neutrino sector have implications
concerning antimatter gravitational asymmetry.  

Gasperini\cite{gasperini} pointed out, in 1989, 
that if antineutrinos and their
corresponding neutrinos have different gravitational coupling
(or if the mass to gravitational force ratio differs between flavors of
neutrinos),
this leads to the local gravitational field contributing to the
transition probability between two different flavor eigenstates.
At the time, no experiments had yet observed neutrino oscillations, 
but the limits which were established were within an order of magnitude of
what since has been observed in accelerator neutrino experiments.
The analysis of such experiments combines the familiar flavor-mixing
angles and mass differences, with a parameter characterizing the antimatter
gravitational asymmetry $\Delta \alpha$. 
A fit of the data sets an upper bound $\Delta \alpha < 0.2$ if the 
gravitational potential is taken to be that of the Earth.  
However, if the galactic gravitational potential is used, the bound 
becomes 0.02\%.  

Minakata and Nunokawa\cite{minakata} study the effects of 
gravitational asymmetry spoiling the (now-accepted) oscillation
mechanism explaining the solar neutrino ``deficit.''  
They place a limit on the possible asymmetry at the same level as
does Gasperini, if they also use the galactic gravitational potential
as their splitting field. 

LoSecco\cite{losecco} studies supernova SN1987A, and
demonstrates that neutrinos and antineutrinos arrived at
(within statistics) the same time (and at the same time as the 
photons).  This implies, following the reasoning of Krauss and Tremaine
\cite{krauss}, that the antineutrino gravitational asymmetry is less than
0.5\%.  Again, the galactic gravitational potential plays a large role in 
this reasoning.

This category of observations shares the same questionable characteristic 
with the time-scaling experiments:  a critical dependence on the use of an
absolute gravitational potential.  
Moreover, the limits placed on gravitational 
asymmetry are much less stringent, and since these results are restricted
to the neutrino sector, any hypothetical asymmetry due to differences in
baryon and antibaryon couplings are not addressed.  For these reasons,
the neutrino-based observations are ``dominated'' by the cyclotron-frequency
and kaon arguments, and need not be a concern when asking whether an effect
is ruled out at the 1\% level.

\section{Theoretical Obstacles to Gravitational Asymmetry}
\label{obstacles}

Several potential theoretical obstacles to $\bar{g} \neq g$ come to mind:

There might be difficulty formulating a framework in which $\bar{g} \neq g$.
However, in section 5 of Nieto and Goldberg\cite{nieto} such frameworks
are presented.

It might seem that CPT invariance forces $\bar{g} = g$.  However, 
in the context of a non-asymptotically flat space-time, or of a
possible underlying
theory which may not exhibit locality (e.g., string theory), CPT invariance
is not sacrosanct, and in fact is a principle which can and should be
tested.  (Indeed, many CPT tests have been done or are in progress.)
The proposed experiment can be viewed as such a test.  Chardin
and Rax\cite{chardin} express doubt that the CPT theorem can be
demonstrated without modification for gravitation.

Various forms of the equivalence principle (see Will's book\cite{will} for
an extensive collection) might demand that $\bar{g} = g$.  
Again, absent independent proof of the WEP, that only transforms the proposed 
experiment into a test of the Weak Equivalence Principle in a realm previously 
unexplored.

The most compelling theoretical objection is in the form of a gedanken
experiment first discussed by Morrison\cite{morrison}.  
Say one starts with an $e^+e^-$ pair starting at height $h_0$, travelling 
upward in a gravitational
field with just enough kinetic energy to reach height $h_1 > h_0$, 
and allows the pair to annihilate, producing two photons, at $h_1$.   
Suppose those photons are directed, by heavy and perfectly reflecting mirrors, 
such that they meet again at $h_0$, 
and that 
they photoproduce an $e^+e^-$ pair at the original height $h_0$.
Those photons will have gained energy by blueshift as they descend,
so that the newly produced pair will have some 
computable kinetic energy, which, assuming the Equivalence Principle and
symmetric gravitation, will be the same as the initial kinetic energy.
If the positron (or the electron) were ``deficient'' from the viewpoint of 
gravitational attraction to the Earth, then the total 
energy gained by the photons as
they descend would be greater than the sum of the kinetic energies 
needed to raise the
pair to the specified height.  Thus an energy non-conservation paradox
would appear, forbidding the gravitational asymmetry in the first place.

Nieto and Goldman\cite{nieto}, in chapter 5, rethink this analysis, 
using the coupled equations including
the Einsteinian (gravitational) field.  
They conclude that in pure tensor gravity, the Morrison argument goes
through, and this forbids ``antigravity.'' But they then show how 
gravivector and graviscalar fields can couple to lepton or baryon number
without leading to a violation of energy conservation. 

Chardin and Rax\cite{chardin}, on the other hand, carefully analyze the
gedanken experiment even with only tensor gravity.  
They note that the microscopic
version of the effect induces a vacuum instability in the presence of a 
gravitational field.  However, they do not take this to exclude antigravity!
They point
out that the typical energy extractable from such a photon at the surface
of the Earth would correspond to a wavelength of a light year, and that
this vacuum instability is similar to the Hawking radiation 
effect which introduces time-asymmetry to general relativity.
``Antigravity is just the tidal effect on the vacuum needed
to induce the temperature suggested by the `naive' expression of the
equivalence principle...''.  
This is certainly no proof that gravitational asymmetry should exist,
but it at least refutes the idea that the Morrison gedanken experiment 
forbids gravitational asymmetry.

\section{Conclusions}

We see from the above that the value of $\bar{g}/g$ has not been measured
by direct observation, 
nor has a value other than one been irrefutably ruled out by theoretical
arguments.
Indirect evidence does suggest, however, that
the value is one to a much greater precision than the 1\% level,
$10^{-4}$ or  perhaps as small as $10^{-8}$.  
(The proposed 1\% measurement can be the first step toward much more 
precise measurements; a measurement to one part in $10^{10}$ appears possible.
But the question at hand addresses the 1\% level.)

Even assuming that one accepts the premise that inequality of $\bar{g}$ and $g$
(at a level of 1\%) has been ruled out  
inferentially (by the time-scaling effects discussed above),
no result directly precludes
a 1\% effect in the antihydrogen experiment.
The issue then becomes whether the value of finally
performing a direct observation is worth the expense and effort of the
experiment.  The proposers feel that it is.  

To do a study of long-range violations of General Relativity in situations
involving antimatter, 
one must make multiple observations under different circumstances; 
the key point would be agreement or discrepancy between the results. 
This is the same strategy proposed for the (billion dollar)
Stage IV dark energy program\cite{detf}, where multiple observations sensitive
to dark energy will be compared for possible discrepancies that
could, for instance, signal a problem with General Relativity on cosmological
scales.

In our case,
the cyclotron frequency of the antiproton is one basic observation,
with length scale of the distance to the local strong galactic attractor.
The direct measurement of antihydrogen deflection in the Earth's gravitational
field would constitute another basic observation, on a very different length scale.
Even a negative result would then become part of a true suite of 
experimental tests 
of the range of validity of General Relativity.

\end{document}